\documentclass[twoside,12pt]{article}
\usepackage{epsfig}

\newcommand{\be}{\begin{equation}}
\newcommand{\ee}{\end{equation}}
\newcommand{\bea}{\begin{eqnarray}}
\newcommand{\eea}{\end{eqnarray}}

\begin{document}

\title{ \vspace{1cm} Alpha decay chains from superheavy nuclei}
\author{C.\ Samanta$^{1,2,3}$\\
\\
$^1$Physics Department, University of Richmond, Richmond, VA 23173, USA\\
$^2$Saha Institute of Nuclear Physics, 1/AF, Bidhannagar, Kolkata-700064, India\\
$^3$Physics Department, Virginia Commonwealth University, Richmond,\\
VA 23284-2000, U.S.A.\\}

\maketitle
\begin{abstract} 

Magic islands for extra-stable nuclei in the midst of the sea of fission-instability were predicted to be around Z=114, 124 or, 126 with N=184, and Z=120, with N=172. Whether these fission-survived superheavy nuclei with high Z and N would live long enough for detection or, undergo alpha-decay in a very short time remains an open question. Alpha-decay half lives of nuclei with $130 \geq Z \geq 100$ have been calculated in a WKB framework using density-dependent M3Y interaction with Q-values from different mass formulae. The results are in excellent agreement with the experimental data. Fission survived Sg nuclei with Z=106, N=162 is predicted to have the highest alpha-decay half life ($\sim$ 3.2 hrs) in the Z=106-108, N=160-164 region called, small island/peninsula. Superheavy nuclei with Z $>$ 118 are found to have alpha-decay half lives of the order of microseconds or, less.

\end{abstract}
\section{Introduction}

One of the great advances in 20th century atomic science was the development of the microscopic theory of the nucleus. In 1965 Myers and Swiatecki [1] pointed out that the shell corrections added to liquid drop model indicates the possibility of closed shells at Z=114 and N=184 which was later confirmed by A. Sobiczewski, F. A. Gareev, B. N. Kalinkin [2]. In 1969 Nilsson et al. [3] predicted that the longest fission half-life center rather symmetrically around the nucleon numbers Z=114, N=184, and the stability against spontaneous fission in this region is due to extra binding resulting from the shell effect which essentially increases the alpha half-lives for nuclei with Z$<$114 and N$<$184 and decreases those for nuclei with Z$>$114 and N$>$184. In 1969, Mosel and Greiner investigated the dependence of the fission barrier on the level-distributions at the Fermi-surface and calculated $\alpha$-decay half lives [4]. 
 
In 1972, Fiset and Nix [5] predicted that the nucleus Z=110, N=184 has the longest total half-life T $\sim 10^{9.4}$ years which is comparable to the age of the earth ($\sim 4.54 \times 10^9$ years). This nucleus was predicted to decay predominantly by  $\alpha$-emission. According to them, as a general rule, the predominant decay mode is  $\alpha$-emission for nuclei containing more than 110 protons, or a few more neutrons than 184; $\beta$-emission for nuclei containing less than 110 protons; and SF for nuclei containing either less than 184 neutrons or, substantially more. Once the closed proton shell at Z=114 is reached, the effect of single particles on the alpha-decay rate is reversed; $\alpha$ decay probability is enhanced for nuclei decaying towards closed shell, while it is hindered for nuclei decaying away. This causes the predominant decay mode to switch to electron capture at Z=114, N=174. Because of the odd-particle hindrance against $\alpha$ decay and SF, and the odd-particle enhancement of electron capture, the nucleus Z=120, N=181 was predicted to decay predominantly by $\alpha$ decay with $\alpha$ decay half-life $T_\alpha \sim 5.3$~ms that is greater than the value $T_\alpha \sim 0.48$~ms for the nucleus Z=120, N=182. An isomer is an excited quantum-mechanical state of a nucleus, in which a combination of nuclear structure effects inhibits its decay and endows the isomeric state with a lifetime that is longer than expected. In principle, a superheavy nucleus might have a long-lived isomer. 

Theoretical predictions of long-lived superheavy nuclei prompted a world-wide search for such nuclei in nature [6], but none has been found so far. Recently a group under the leadership of A. Marinov [7] has claimed the first evidence for the existence of a superheavy nucleus in nature with atomic mass number A=292 and abundance (1-10)$\times 10^{-12}$ relative to $^{232}$Th in a study of natural Th using inductively coupled plasma-sector field mass spectrometry. It was conjectured that it might belong to a new class of long-lived high-spin super- and hyperdeformed isomeric states of Z=122 or nearby element with estimated half-life of $\sim 10^8$ years, but no definite proof of this element was established.

The shell structure of superheavy nuclei was investigated within various parameterizations of relativistic and non-relativistic nuclear mean-field models and nuclei with (Z =114, N=184), (Z=120, N=172) or, (Z=126, N=184) were found to be doubly-magic [8]. Cwiok, Heenen and W.Nazarewicz [9] predicted that the long-lived superheavy elements can exist in a variety of shapes, including spherical, axial and triaxial configurations. Only when N=184 is approached one expects superheavy nuclei that are spherical in their ground states. 

In the beginning of the 1980's the first observations of the elements with Z=107-109 were made at GSI, Germany [10]. In 1994,  $\alpha$-decay chains were observed from nucleus $^{269}110$ and later on, $\alpha$-decay chains from nuclides $^{271}110$, $^{272}111$, $^{277}112$, $^{283}112$ were detected at GSI [11]. Recently, the doubly-magic deformed $^{270}Hs$ (Z=108, N=162) superheavy nucleus has been produced [12]. In Japan, RIKEN reconfirmed the  $\alpha$-decay chains from $^{271}110$, $^{272}111$ and $^{277}112$ [13 -15]. They also found signature of the superheavy nucleus $^{278}113$. So far, JINR, Russia has reported $\alpha$-decay chains of nuclei with Z=106-116 and 118 [16-19]. None of these experiments has reached the N=184 region as yet. SHE with Z=106, 107, 108, 112 and recently, 114 have been chemically characterized [20].

\section{Calculations}

The $\alpha$-decay half lives have been calculated extensively through different models [21]. Calculations in the framework of quantum mechanical tunneling of an $\alpha$-particle from a parent nucleus has been found to provide an excellent description of the experimental data when experimental Q-values are employed along with density-dependent M3Y interaction [22]. Calculation detail of the $\alpha$-decay half lives of superheavy nuclei in this framework has been described in references [22 - 27]. Only a brief outline of the method [28-29] is given here.
The required nuclear interaction potentials are calculated by double folding the density distribution functions of the  $\alpha$-particle and the daughter nucleus with density-dependent M3Y effective interaction. The microscopic  $\alpha$-nucleus potential thus obtained, along with the Coulomb interaction potential and the minimum centrifugal barrier required for the spin-parity conservation, form the potential barrier. 
The action integral $K$ within the WKB approximation is given by

\begin{equation}
 K = (2/\hbar) \int_{R_a}^{R_b} {[2\mu (E(R) - E_v - Q)]}^{1/2} dR
\label{seqn12}
\end{equation}
\noindent
where the total interaction energy $E(R)$ between the $\alpha$ and the residual daughter nucleus is equal to the sum of the nuclear interaction energy, Coulomb interaction energy and the centrifugal barrier.

\begin{equation}
 E(R) = V_N(R) + V_C(R) + \hbar^2 c^2 l(l+1) / (2\mu R^2)
\label{seqn13}
\end{equation}   
\noindent
where the reduced mass $\mu = M_{\alpha} M_d/ (M_{\alpha} + M_d)$ and $V_C(R)$ is the Coulomb potential between the $\alpha$ and the residual daughter nucleus. $R_a$ and $R_b$ are the second and third turning points of the WKB action integral determined from the equations 

\begin{equation}
 E(R_a)  = Q + E_v =  E(R_b)
\label{seqn14}
\end{equation}
\noindent
whose solutions provide three turning points.

The $\alpha$-particle oscillates between 1st and 2nd turning points and tunnels through the barrier at 2nd and 3rd TP. The zero point vibration energy  $E_v \propto$ Q, where $E_v=0.1045$Q for even Z-even N, 0.0962Q for odd Z-even N, 0.0907Q for even Z-odd N, 0.0767Q for odd -odd  parent nuclei (includes pairing and shell effects) [21]. The decay half life of the parent nucleus is,

\begin{equation}
 T_{1/2} = [(h \ln2) / (2 E_v)] [1 + \exp(K)]
\label{seqn11}
\end{equation}
\noindent
       
The calculated half lives are very sensitive to Q, as it goes to the exponential function in Eq. (4) through the action integral in Eq. (1). Theoretical Q-values are taken from three different mass formulas: (i) Muntian-Hofmann-Patyk-Sobiczewski (Q-MMM) [30], (ii) Myers-Swiatecki (Q-MS) [31] and, (iii) Koura-Uno-Tachibana-Yamada (Q-KUTY) [32]. Fission half lives are taken from the experimental data and the predictions of Smolanczuk et al. [33]. Beta-decay half lives are taken from Moller-Nix-Kratz [34].

\section{Results and discussions}

The $\alpha$-decay half lives of about 1700 isotopes of elements with $100\leq \rm Z \leq 130$ have been calculated [26]. Calculations with Q-values from experiment and Q-MMM well reproduce the experimental data for even-even nuclei with $l=0$ transition [22-28]. For some odd-odd or, odd A nuclei, $l \neq 0$ is needed.  With Q-KUTY, at N=184, the $T_\alpha$ values for Z=110, 112, 114 are $\sim 10^{10}$s, $10^8$s, $10^6$s respectively. Theoretical fission half-life values ($T_{SF}$) of these nuclei are $\sim 10^{12}$s, $10^{13}$s, $10^{13}$s respectively. The nuclei $^{296}112$ (N=184), $^{298}114$ (N=184) are $\beta$-stable whereas, $^{294}110$ (N=184) is predicted to have large $T_\beta$ in ref. [34] due to its very small positive $Q_\beta$-value. 

In the initial RIKEN data for Z=113 in 2004 there was some discrepancy. As Q value decreases, the half life value should generally increase, but an opposite trend was observed in the decay of 111 and 109.  There is no discrepancy in the repeat data of RIKEN (in 2007).

For the $^{277}112$ and its alpha-decay chain, some discrepancies were observed between the GSI and RIKEN data [14]. While the observed first four $\alpha$-decay chains of GSI and RIKEN are similar, the chain-3 of GSI extends up to $^{257}No$. Calculations [25] in a quantum tunneling model, with $l=0$, reasonably reproduce the experimental data of $\alpha_2$ and $\alpha_3$ decay channels. For the $\alpha_1$ decay the experimental data are much higher than the theoretical predictions (with $l=0$), and $l \sim 7$ can explain the data. But, for the $\alpha_4$, $\alpha_5$ and $\alpha_6$ decays, the calculated $\alpha$-decay half lives (even with $l=0$) are higher than the experimental ones which can not be explained. The discrepancy might arise from the fact that the theoretical Q values considered here are for the decay from the ground state of the parent to the ground state of the daughter, although there is no guarantee that the experimentally observed alpha decay chains proceed from the ground state of the parent to the ground state of the daughter. In fact, transitions to and from excited states are also possible [35]. 
In this context, further experimental data with higher statistics would also be useful. \\

Detailed investigation [22-28] indicates that in the region Z =106-108 and N$\sim 160-164$ a small 'island/peninsula' might survive fission and $\beta$-decay, and superheavy nuclei in this region might predominantly undergo $\alpha$-decay. 
A careful scrutiny also reveals that $^{298}114$ is not the center of the magic island as predicted earlier [3]. On the contrary, the nucleus with Z=110, N=183 appears to be near the center of a possible `magic island' (Z=104 -116, N$\sim$176 -186) with $T_\alpha \sim 352$ years (with Q-KUTY). The nucleus $^{290}Sg$ has $T_\alpha\sim 10^8$ years (with Q-KUTY) and $T_{SF}\sim 10^6$ years. Therefore, it might have longer life time compared to other super-heavies. However, for both $^{293}Ds$ (N=183) and $^{290}Sg$ (N=184) nuclei, $\beta$-decay might be another possible decay mode with large $T_\beta$ values.\\

\begin{table}
\begin{center}
\caption{\label{table1} $\alpha$-decay half lives of Z=118 chain [16-19] and Z=112} 
\begin{tabular}{r|r|r|r|r}
\hline
Parent    &Expt.                      &Theory(MS)           &Expt.                 & This work   \\ 
Nuclei $^AZ$    &Q (MeV)& Q (MeV) &$T_{1/2}$ &$T_{1/2}$ \\  

 \hline
&&&&\\

$^{294}118$&$11.81 \pm 0.06$&$12.51$&$0.89^{+1.07}_{-0.31}$ ms &$0.66^{+0.23}_{-0.18}$ ms \\ 
&&&&\\

$^{290}116$&$11.00 \pm 0.08$& 11.34 & $15.0^{+26}_{-6.0}$ ms & $13.4^{+7.7}_{-5.2} $ms \\ 
&&&&\\

$^{286}114$&$10.35 \pm 0.06$ &9.61&$0.16^{+0.07}_{-0.03}$ s& $0.14^{+0.06}_{-0.04}$s\\ 
&&&&\\

$^{282}112$ & SF&  &  &  \\
&&&&\\ 
$^{283}112$&$9.67 \pm 0.06$ &9.22&$4.0^{+1.3}_{-0.7}$ s& $5.9^{+2.9}_{-2.0}$s\\ 
&&&&\\
$^{279}110$&$9.84 \pm 0.06$ &9.89&$0.18^{+0.05}_{-0.03}$ s& $0.40^{+0.18}_{-0.13}$s\\ 
&&&&\\
$^{275}108$&$9.44 \pm 0.07$ &9.58&$0.15^{+0.27}_{-0.06}$ s& $1.09^{+0.73}_{-0.40}$s\\ 
&&&&\\
$^{271}106$&$8.65 \pm 0.08$ &8.59&$2.40^{+4.3}_{-1.0}$ min& $1.0^{+0.8}_{-0.50}$min\\ 
&&&&\\
$^{267}104$ & SF&  &  &  \\
\hline
\end{tabular} 
\end{center}
\end{table}

\begin{figure}[tb]
\begin{center}
\epsfig{file=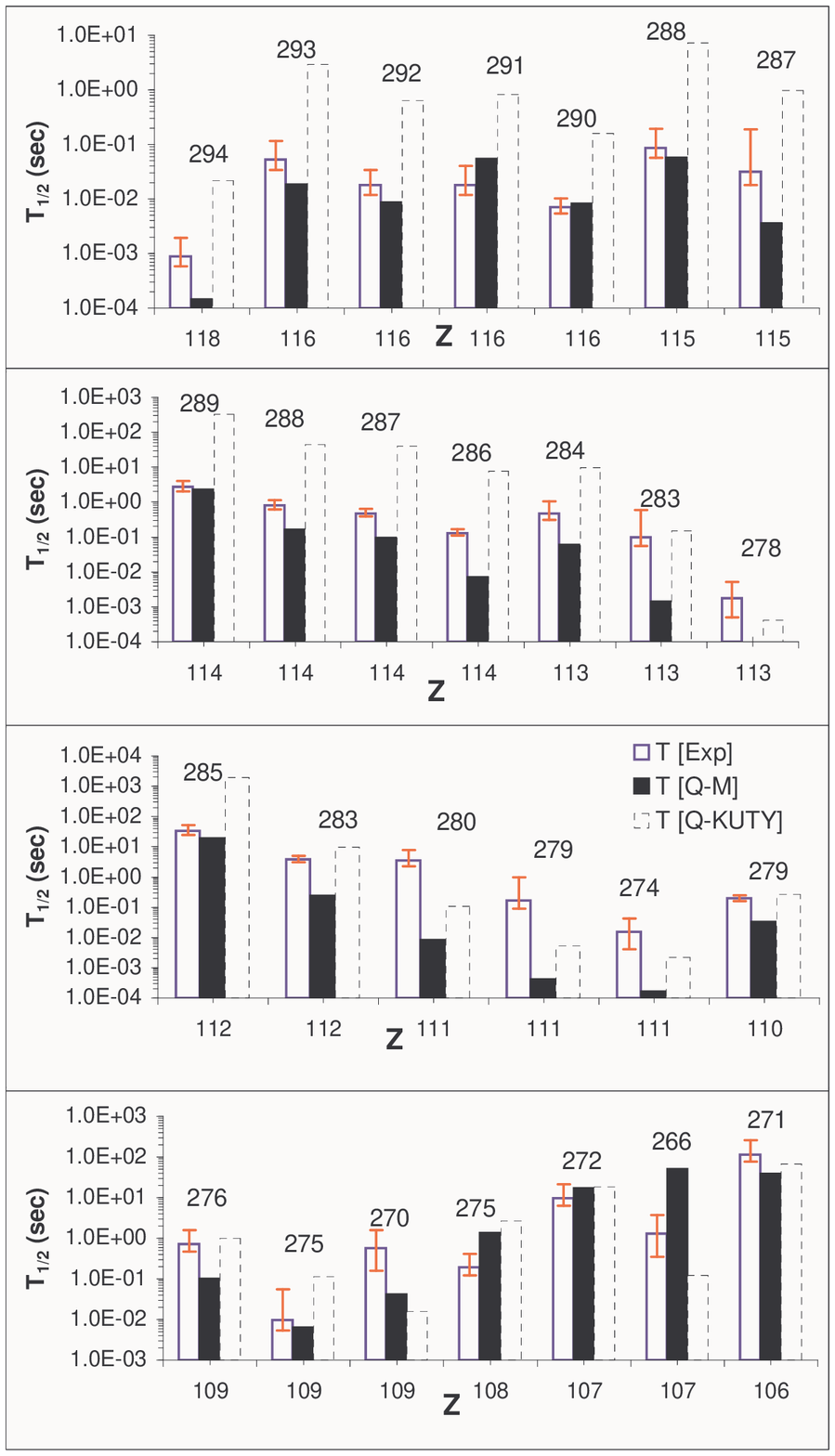,scale=0.70}
\begin{minipage}[t]{10.5 cm}
\caption
{Plots of $\alpha$-decay half life [T$_{1/2}$(sec)] \cite{er26} versus proton number Z for different mass number A (indicated  on top of each coloumn). (a) Hollow columns of solid lines with error bars are experimental $\alpha$-decay half lives (T~[Exp]), (b) filled columns are theoretical half lives ($T~[Q$-$M]$=$T_{1/2}[Q^{M}_{th}]$) in WKB frame work with DDM3Y interaction and [$Q^{M}_{th}$] from Muntian-Patyk-Hofmann-Sobiczewski mass formula, (c) hollow columns of dashed lines are ($T~[Q$-$KUTY]$=$T_{1/2}[Q^{KUTY}_{th}]$) in the same framework but with [$Q^{KUTY}_{th}$] from Koura-Tachibana-Uno-Yamada mass estimates.}
\end{minipage}
\end{center}
\end{figure}

\section{Summary} 

In mid-sixties it was predicted that beyond actinides (Z = 89-103), there exists a region called, "Magic Island", or, "Island of stability" at $320 \geq A \geq 250$  where one can find superheavy elements (SHE) with large life times. This prompted a world-wide search for such nuclei both in nature as well as in the laboratories. While several SHE have been produced in the laboratories none has been found in nature so far. In the early 1990s Peter Armbruster, Sigurd Hofmann, Gottfried M\"unzenberg and co-workers at the GSI laboratory in Darmstadt, Germany, used cold-fusion reactions to synthesize elements 107-112. These data were later confirmed by Kosuke Morita and co-workers at the RIKEN, Japan who also synthesized elements 110 and 111, 112, 113 in cold-fusion reactions. Isotopes of the SHE 112, 113, 114, 115, 116 and the element  A = 294, Z = 118 have been produced in fusion evaporation reactions at Flerov Laboratory of Nuclear Reactions (FLNR)-Joint Institute for Nuclear Research (JINR), Dubna, Russia by Yu.Ts. Oganessian and co-workers. SHE with Z = 106, 107, 108, 112 and recently, 114 have been chemically characterized. Elements above 111 have not  been named so far.

We calculated the alpha-decay half-lives of SHE with $130 \geq Z \geq 100$ in a quantum tunneling model with density-dependent M3Y interaction. Theoretical Q-values are taken from three different mass formulae. The experimental data are in excellent agreement with theoretical calculations. In the N=184 region, use of Q-KUTY in above formalism indicates some long-lived superheavy nuclei. For example, $^{290}$Sg (Z=106) has $T_\alpha \sim 10^8$ years (although $\beta$-decay of this nucleus might be possible). Contrary to earlier prediction [5], the nucleus $^{294}$Ds (Z=110) has $T_\alpha \sim 311$ years, a value much less than the age of the earth. Alpha-decay half lives of $^{294}$110, $^{296}$112, $^{298}$114 are $\sim$ 311 years, 3.10 years and 17 days respectively. These are very neutron rich nuclei and it is difficult to produce such superheavy nuclei with the existing facilities. However, with the upcoming RIB facilities and improved detection technique, we believe that such long-lived SHE would be synthesized in future. Overall estimation suggests that the nucleus with Z = 110, N = 183 will be near the center of a magic island. For superheavy nuclei with Z $\geq$ 116 and N $\sim$ 184 the alpha-decay half-lives are less than one second. In fact, Z=120, 124, 126 with N=184 might form spherical doubly-magic nuclei and survive fission but, they would undergo alpha-decay within microseconds or, less. The scenario changes when there are long-lived isomeric states in heavy nuclei.
Two isomeric states have been found in $^{254}$No, with 102 protons and 152 neutrons [36]. Possibility of long-lived isomeric state in heavier nuclei needs careful investigation.   \\  
   
{\bf Acknowledgments}\\

It is a pleasure to thank D.N. Basu and P. Roy Chowdhury without whom this work would not have been possible. The author gratefully acknowledges the funding provided by the Saha Institute of Nuclear Physics, Kolkata, India and the University of Richmond, Richmond, Virginia, USA.


\begin{thebibliography}{99}
\itemsep -2pt 
\bibitem{er1} W.D. Myers and W.J. Swiatecki, {\it Report UCRL} (1965) 11980.
\bibitem{er2} A. Sobiczewski, F. A. Gareev, B.N. Kalinkin, {\it Phys. Lett.} 22 (1966) 500.
\bibitem{er3} S.G.Nilsson, C.F. Tsang, A. Sobiczes, Z. Szymansk, S. Wycech, C. Gustafso, I.L. Lamm, P. Moeller, B. Nilsson {\it Nucl. Phys.} A 131 (1969) 1.
\bibitem{er4} U. Mosel, W. Greiner, {\it Z. Phys.} 222 (1969) 261; W. Greiner, {\it Int. J. Mod. Phys.} E 5 (1995) 1.
\bibitem{er5} E. O. Fiset and J. R. Nix, {it Nucl. Phys.} A 193 (1972) 647.
\bibitem{er6} R. K. Bull, {\it Nature} 282 (1979) 393; T. Lund, R. Brandt, D. Molzahn, G. Tress, P. Vater and A. Marinov, {\it Z. für Phys.} A 300 (1981) 285; K. Murtazaev, V. P.  Perelygin, {\it Sov. At. Energy} 63 ( 1987) 407.
\bibitem{er7} A. Marinov et al., arXiv:0804.3869v1 [nucl-ex].
\bibitem{er8} A. T. Kruppa, M. Bender, W. Nazarewicz, P.-G. Reinhard, T. Vertse, and S. Cwiok, {\it Phys. Rev.} C 61 (2000) 034313.
\bibitem{er9} S.  Cwiok, P.-H. Heenen and W. Nazarewicz, {\it Nature} 433 (2005) 705; R.-D. Herzberg et al., {\it Nature} 442 (2006) 896.
\bibitem{er10} P. Armbruster, {\it Acta Phys. Pol.} B 34 (2003) 1825; P. Armbruster, {\it Annu. Rev. Nucl. Part. Sci.} 35 (1985) 135.
\bibitem{er11} S. Hofmann et al., {\it Euro. Phys. Jour.} A 32 (2007) 251; S. Hofmann, G. Munzenberg, {\it Rev. Mod. Phys.} 72 (2000) 733; S. Hofmann, {\it Rep. Prog. Phys.} 61 (1998) 639; G. Munzenberg, {\it Rep. Prog. Phys.} 51 (1988) 57.
\bibitem{er12} J. Dvorak  et al., {\it Phys. Rev. Lett.}  97 (2006) 242501. 
\bibitem{er13} K. Morita et al., {\it J. Phys. Soc. Jpn.} 73 (2004) 2593.
\bibitem{er14} K. Morita et al., {\it J. Phys. Soc. Jpn.} 76 (2007) 045001; (2007) 043201.
\bibitem{er15} K. Morita et al., {\it J. Phys. Soc. Jpn.} 73 (2004) 1738.
\bibitem{er16} Yu.Ts. Oganessian, {\it Euro. Phys. Jour.} D 45 (2007) 17; {\it Phys. Rev.} C 76 (2007) 011601(R); {\it J. Phys.} G 34 (2007) R165.
\bibitem{er17} Yu.Ts. Oganessian et al., {\it Phys. Rev.} C 74 (2006) 044602.
\bibitem{er18} Yu.Ts. Oganessian et al., {\it Phys. Rev.} C 72 (2005) 034611; {\it Phys. Rev.} C 71 (2005) 029902(E). 
\bibitem{er19} Yu.Ts. Oganessian et al., {\it Phys. Rev.} C 70 (2004) 064609; {\it Phys. Rev.} C 69 (2004) 021601(R).
\bibitem{er20} R. Eichler et al., {\it Nature} 447 (2007) 72; Andreas Türler, {\it Nature} 447 (2007) 47; S.N. Dmitriev, NRC7- Seventh International Conference on Nuclear and Radiochemistry, Budapest, Hungary 24-29 August 2008. 
\bibitem{er21} D.N. Poenaru, W. Greiner, K. Depta, M. Ivascu, D. Mazilu and A. Sandulescu, {\it Atomic Data and Nuclear Data Tables} 34 (1986) 423-538; D.N. Poenaru, I-H. Plonski, and Walter Greiner, {\it Phys. Rev.} C 74 (2006) 014312; D.N. Poenaru, I-H. Plonski, R.A. Gherghescu and Walter Greiner, {\it J. Phys. G: Nucl. Part. Phys.} 32 (2006) 1223.
\bibitem{er22} P. Roy Chowdhury, C. Samanta and D.N. Basu, {\it Phys. Rev.} C 73 (2006) 014612. 
\bibitem{er23} C. Samanta, P. Roy Chowdhury and D.N. Basu, {\it Nucl. Phys.} A 789 (2007) 142. 
\bibitem{er24} P. Roy Chowdhury, D.N. Basu  and C. Samanta, {\it Phys. Rev.} C 75 (2007) 047306.
\bibitem{er25} C. Samanta, D.N. Basu and P. Roy Chowdhury, {\it J. Phys. Soc. Jpn.} 76 (2007) 124201.
\bibitem{er26} P. Roy Chowdhury, C. Samanta and D.N. Basu, {\it Atomic Data and Nuclear Data Tables} (2008) 94 (2008) 781. 
\bibitem{er27} P. Roy Chowdhury, C. Samanta and D.N. Basu, {\it Phys. Rev.} C 77 (2008) 044603.
\bibitem{er28} C. Samanta, Rom. Rep. Phys.59, 491 (2007).
\bibitem{er29} C. Samanta, Procceedings of the Fourth International conference on Fission and properties of neutron-rich nuclei,  Sanibel Island, 11-17 November 2007, World Scientific, Ed. by J.H. Hamilton, A.V. Ramayya, H.K. Carter, p. 329.
\bibitem{er30} I. Muntian, Z. Patyk and A. Sobiczewski, {\it Acta Phys. Pol.} B 32 (2001) 691; I. Muntian, S. Hofmann, Z. Patyk and A. Sobiczewski, {\it Acta Phys. Pol.} B 34 (2003) 2073; I. Muntian, Z. Patyk and A. Sobiczewski, {\it Phys. At. Nucl.} 66 (2003) 1015.
\bibitem{er31} W.D. Myers and W.J. Swiatecki, Lawrence Berkeley Laboratory preprint LBL-36803, Dec. (1994); {\it Nucl. Phys.} A 601 (1996) 141.
\bibitem{er32} H. Koura, T. Tachibana, M. Uno and M. Yamada, KUTY mass formula 2005 revised version, {\it Prog. In Theor. Phys.} 113 (2005) 305.
\bibitem{er33} R. Smolanczuk, J. Skalski, and A. Sobiczewski, {\it Phys. Rev.} C 52 (1995) 1871; R. Smolanczuk {\it Phys. Rev.} C 56 (1997) 812.
\bibitem{er34} P. Moller, J. R. Nix, and K.-L. Kratz, {\it Atomic Data Nuclear Data Tables} 66 (1997) 131.
\bibitem{er35} D.S. Delion, R.J. Liotta, R. Wyss, {\it Phys. Rev.} C 76 (2007) 044301.
\bibitem{er36} R.-D. Herzberg  et al., Nature 442 (2006) 896. 

\end{thebibliography}
\end{document}